\documentclass[twocolumn,showpacs,preprintnumbers,amsmath,amssymb]{revtex4}

\usepackage{graphicx}	
\graphicspath{{figcondmat/}}
\usepackage{dcolumn}	
\usepackage{bm,amsfonts,stmaryrd,wasysym,pifont,textcomp}
\usepackage[mathcal]{euscript}
\usepackage{enumerate}
\usepackage{epsfig}
\usepackage{color}

\definecolor{bleu}{rgb}{0,0,1}
\definecolor{vert}{rgb}{0,0.5,0}
\definecolor{rouge}{rgb}{1,0,0}
\definecolor{rose}{rgb}{0.9,0.3,0.7}
\definecolor{azur}{rgb}{0,0.5,0.5}
\definecolor{orange}{rgb}{1,0.5,0.2}

\newcommand{\be}{\begin{equation}}
\newcommand{\ee}{\end{equation}}
\newcommand{\ba}{\begin{eqnarray}}
\newcommand{\ea}{\end{eqnarray}}

\newcommand{\ie}{\textit{i.e.} }

\newcommand{\leg}[1]{\textbf{#1}}
\newcommand{\inset}{\textit{\textbf{Inset}} }

\begin{document}

\title{Journey of an intruder through the fluidisation and jamming transitions\\of a dense granular media}

\author{Rapha\"el Candelier}
 \email{raphael.candelier@cea.fr}
 \affiliation{GIT-SPEC, CEA-Saclay, 91 191 Gif-sur-Yvette, France}
\author{Olivier Dauchot}
 \email{olivier.dauchot@cea.fr}
 \affiliation{GIT-SPEC, CEA-Saclay, 91 191 Gif-sur-Yvette, France}
 \homepage{http://iramis.cea.fr/spec/GIT/}

\date{\today}

\begin{abstract}
We study experimentally the motion of an intruder dragged into an amorphous monolayer of horizontally vibrated grains at high packing fractions. This motion exhibits two transitions. The first transition separates a continuous motion regime at comparatively low packing fractions and large dragging force from an intermittent motion one at high packing fraction and low dragging force. Associated to these different motions, we observe a transition from a linear rheology to a stiffer response. We thereby call ``fluidisation'' this first transition. A second transition is observed {\it within} the intermittent regime, when the intruder's motion is made of intermittent bursts separated by long waiting times. We observe a peak in the relative fluctuations of the intruder's displacements and a critical scaling of the burst amplitudes distributions. This transition occurs at the jamming point $\phi_J$ characterized in a previous study~\cite{lechenault2008csa} and defined as the point where the static pressure (\ie the pressure measured in the absence of vibration) vanishes. Investigating the motion of the surrounding grains, we show that below the fluidisation transition, there is a permanent wake of free volume behind the intruder. This transition is marked by the evolution of the reorganization patterns around the intruder, which evolve from compact aggregates in the flowing regime to long-range branched shapes in the intermittent regime, suggesting an increasing role of the stress fluctuations. Remarkably, the distributions of the kinetic energy of these reorganization patterns also exhibits a critical scaling at the jamming transition.
\end{abstract}

\pacs{PACS number}

\maketitle

Jamming occurs, when a system develops a yield stress in a disordered state~\cite{liu1998jnc,ohern2002rpf,ohern2003jzt} and has been reported in a wide class of systems such as colloids~\cite{habdas2004fmp}, foams~\cite{khan1988fri}, emulsions~\cite{brujic2006fcm}, granular materials~\cite{midi2004dgf,majmudar2007tjt} as well as in various model situations ~\cite{hastings2003dfg,ellenbroek2006csl,ohern2002rpf,xu2006mys}.
One possible mechanism for such a change between fluid and solid like behavior is that rearrangements of particles become progressively slower while the stress relaxation time grows dramatically.
The dynamics becomes spatially heterogeneous and temporally intermittent, while the stress response appears more and more heterogeneous. A stringent manifestation of such inhomogeneities is the ``stick-slip'' response observed when the system is driven close to yielding and flows in rapid bursts.
However the interplay between density fluctuations and stress relaxation is still poorly understood 
Questions of interest are as follow: What is the nature of the re-arrangement events? How do these events depend on the external load and packing fraction? 
Also from a more fundamental viewpoint, whether the emergence of a yield stress coincides with dynamical arrest is still a matter of debate. 

Microrheology is a promising technique providing local probes of the dynamics in complex fluid~\cite{waigh2005mcf}. Studying the motion of an intruder embedded in the material of interest, one is able to investigate the microscopic origins of the complex-fluid behavior and in particular the link between microscopic mechanisms and macroscopic properties as given by conventional rheology. Applying a force to the intruder, one explores the non-equilibrium and usually non-linear response, providing detailed insight into the structure-dynamics relationship. Previous drag experiments in colloids~\cite{habdas2004fmp}, foams~\cite{dollet2005tdf}, static~\cite{albert1999sdi} and shaken~\cite{zik1992msv} granular media as well as simulations of structural glasses~\cite{hastings2003dfg} were focusing on the velocity dependence of the drag force: proportionality is found for loose enough systems, reminding Stoke's law, while an increasing yield stress appears for denser packing. Stress fluctuations have been studied in details in~\cite{geng2005sdi}, and spatial reorganizations in ~\cite{kolb2004rdg}, however a clear picture filling the gap between spatial fluctuations and rheological observations is still lacking.

In the present paper, we investigate the motion of an intruder dragged with a constant force, within an amorphous monolayer of horizontally vibrated grains, a system for which the jamming transition has been clearly identified and characterized in terms of the critical behavior of the dynamics in a previous study~\cite{lechenault2008csa}. At moderate packing fractions, and comparatively high force, the intruder moves rapidly as soon as the force is applied. Above some threshold value of the packing fraction which increases with the applied force, the intruder exhibits an intermittent creep motion with strong fluctuations reminiscent of a ``crackling noise'' signal. Simultaneously, the force-velocity relation evolves from a linear rheology to a stiffer response, thereby suggesting to call ``fluidisation'' this first transition.  A second transition is observed {\it within} the intermittent regime, when the intruder's motion is made of intermittent bursts separated by long waiting times. This transition is signed by a peak in the relative fluctuations of the intruder's displacements and a critical scaling of the bursts amplitudes' distributions. This transition occurs at the jamming point $\phi_J$ characterized in~\cite{lechenault2008csa} and defined as the point where the static pressure (\ie the pressure measured when the vibration is switched off) vanishes. 
In~\cite{lechenault2008csa}, the authors demonstrate that dynamical heterogeneities become critical at the transition. Here we investigate the motion of the grains surrounding the intruder. Below the fluidization transition a wake of free volume is observed behind the intruder and the fluidisation transition is marked by the evolution of the reorganization patterns around the intruder, going from compact aggregates in the flowing regime to long-range branched shapes in the intermittent regime, suggesting an increasing role of the stress fluctuations. The distributions of the kinetic energy of these reorganization patterns also exhibits a critical scaling at the jamming transition. 

This paper is an extended version of a recently published letter~\cite{candelier2009cmi}, in which both transitions have been reported. The purpose of the present paper is to provide a comprehensive study of the displacement fields surrounding the intruder and to take this opportunity to provide details on our analysis procedures as well as an extended discussion of our results.  The paper organizes as follows : the experimental set-up and protocols are described in section~\ref{sec:Esap}. In section~\ref{sec:Faj} we first introduce the raw dynamical quantities and the phase diagram (\ref{sec:Pd}), then we characterize the fluidisation (\ref{sec:Fluidisation}) and jamming (\ref{sec:Jamming}) transitions. The dynamics around the intruder is analyzed in section~\ref{sec:Ati}, both by the relation between the average flow and the spatial fluctuations (\ref{sec:Afasf}) and by the evolution of the averaged free volume around the intruder (\ref{sec:Fv}). Finally, we study the bursts statistics in the intermittent regime close to jamming in section~\ref{sec:Bsctj}. A general discussion and a few concluding remarks are given in section~\ref{sec:Dacr}.

\section{Experimental set-up and protocol}
\label{sec:Esap}

\begin{figure}[t] 
\center
\includegraphics[width=0.99\columnwidth]{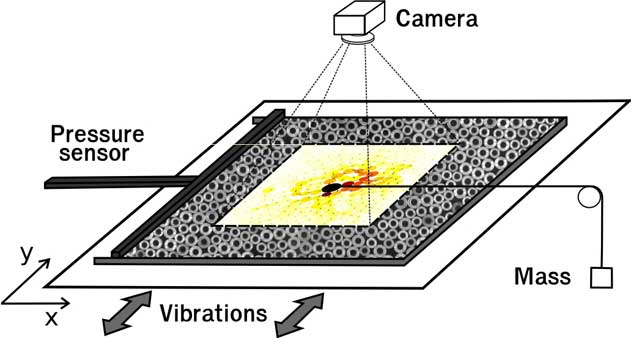}
\includegraphics[width=0.49\columnwidth]{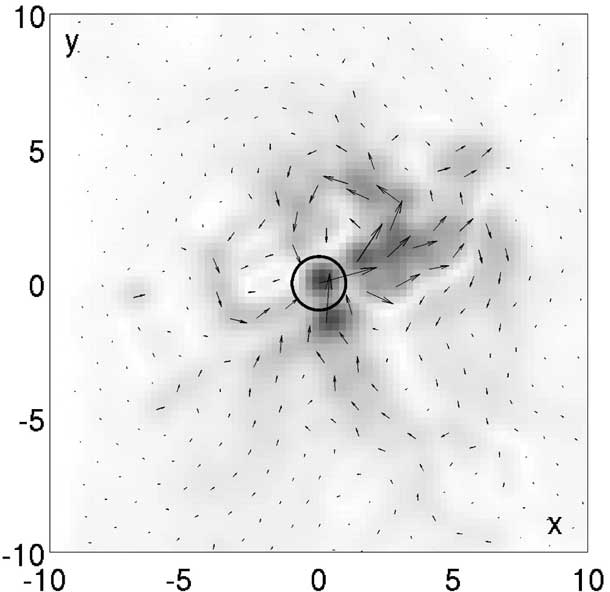}
\includegraphics[width=0.49\columnwidth]{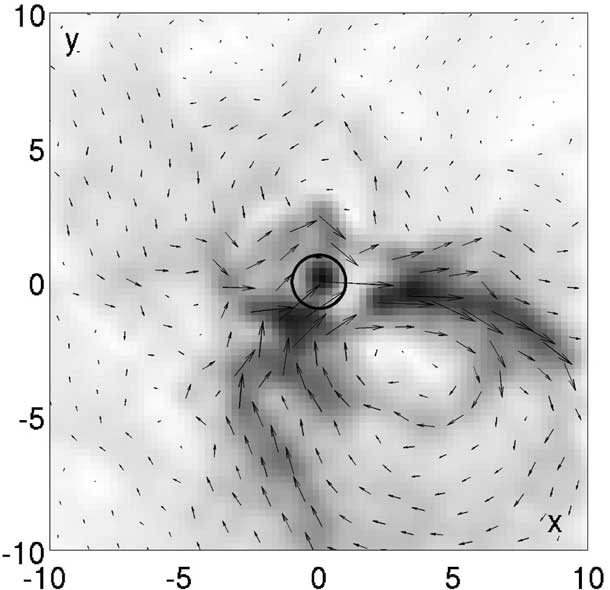}
\caption{
\leg{Top} Experimental Setup (see text for details).
\leg{Bottom} Two samples of displacement fields around the intruder at $\phi_J$ for similar $x$-displacements of the intruder ($\Delta x = 2.14$ and $2.08$, in respectively $14$ and $131$ time steps). The arrows show the unmagnified grain displacements, while the interpolated amplitude field is in grey scale.
}
\label{fig:setup}
\vspace{-0.0cm}
\end{figure}

The experimental set-up has been described elsewhere~\cite{lechenault2008csa} and we shall only recall here its most important characteristics and the modifications induced by the dragging procedure. The system is made of a monolayer of $8500$ bi-disperse brass cylinders of diameters $d_{small} = 4\pm0.01{\rm mm}$ and $d_{big} = 5\pm0.01{\rm mm}$ laid out on a horizontal glass plate vibrated in its plane at a frequency of $10{\rm Hz}$ and with a peak-to-peak amplitude of $10{\rm mm}$. The grains are confined in a cell, fixed in the laboratory frame, the volume of which can be adjusted by a lateral mobile wall controlled by a $\mu$m accuracy translation platen, which allows us to vary the packing fraction $\phi$ of the grains by tiny amounts ($\delta\phi/\phi \sim 5\times10^{-4}$). The pressure exerted on this wall is measured by a force sensor (see fig.\ref{fig:setup}). The intruder consists in a larger particle ($d_{intruder}=2.d_{small}$) of same height introduced in the system and pulled by a mass via a pulley perpendicularly to the vibration. In all data presented here the resultant motion is strongly overdamped and the applied force can be considered as constant. We use a fishing wire that stands over the other grains and doesn't disturb their dynamics. The time unit is set to one plate oscillation while the length unit is chosen to be the diameter of the small particles. The drag forces $F$ are expressed as the ratio of the applied mass onto the total mass of grains in the cell ($M_{total} = 2.365 kg$).

\begin{figure}[t!] 
\center
\includegraphics[width=0.99\columnwidth]{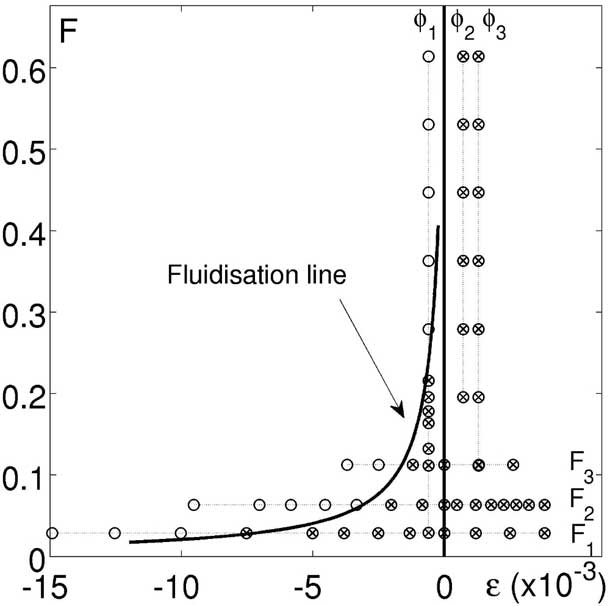}
\caption{
Experimental conditions in a force - reduced packing fraction $\varepsilon$ diagram. Each point corresponds to a trajectory of the intruder into the media.
The horizontal and vertical dotted lines correspond to the experimental exploration paths, either at constant force of packing fraction. $\ocircle$ (resp. $\varotimes$) denote the fluidized (resp. intermittent) behavior (see text for definition).
}
\label{fig:sketch}
\vspace{-0.0cm}
\end{figure}

We have been using two protocols, setting either the drag force or the packing fraction constant. We explore the force~/ packing fraction diagram along constant force lines using three different drag forces ($F_1=0.029$, $F_2=0.064$ and $F_3=0.113$) and varying the packing fractions from $\phi=0.8223$ to $0.8418$, and along constant packing fraction lines using three different packing fractions ($\phi_1=0.8383$, $\phi_2=0.8304$ and $\phi_3=0.8399$) and varying the drag force from $F=0.029$ to $0.617$, as sketched in fig.\ref{fig:sketch}.

Starting from a low packing fraction $\phi$, we gradually compress the system until it reaches a highly jammed state following the same protocol as in~\cite{lechenault2008csa}. Then we stepwise decrease the packing fraction. In the absence of the intruder, it was shown on the one hand that the average relaxation time increases monotonically with the packing fraction and on the other hand that the dynamics exhibits strong dynamical heterogeneities, the length-scale and time-scale of which exhibit a sharp peak at an intermediate packing fraction. The authors have shown that the spatial correlations of these dynamical heterogeneities exhibits a critical scaling at the transition. The pressure measured at the wall in the absence of vibration falls to zero precisely below that packing fraction, hence called the jamming transition $\phi_J$. It is important to mention that for such high packing fractions, the structure as given by the neighborhood relation among the grains is frozen on experimental time-scales. Hence the observed transition is to be understood as the jamming of a given frozen configuration.  Accordingly the value of $\phi_J$ may slightly change from one run to another, since it may differ from one frozen configuration to another. In the present case, by monitoring the pressure at the wall while interrupting the vibration and without drag force, we could localize the jamming transition and obtain three close but different values $\phi_J$~: $0.8369$ ($F_1$), $0.8383$ ($F_2$), $0.8379$($F_3$) and $0.8388$ in the run where we explore iso-$\phi$ lines. These values also are slightly smaller than the value $\phi_J=0.8417$ reported in~\cite{lechenault2008csa}, maybe an effect of the geometrical distortion induced locally by the size of the intruder, twice larger than the other grains. In the following we will use either the packing fraction $\phi$ or the reduced packing fraction $\varepsilon = (\phi-\phi_J)/\phi_J$.

In the present study, before each step, the intruder is removed from the position it has reached, replaced by one big and two small grains, and inserted at its initial position in place of one big and two small grains. Then a downward step in packing fraction is eventually done, and the system is kept under vibration up to one hour in order to "equilibrate" the configuration. At that point, the pressure has the same value as without the intruder for the corresponding packing fraction, indicating that the system has recovered from the small perturbation induced by the intruder's ``teletransportation''. Only then the force is applied and the intruder is dragged through the cell, while its stroboscopic motion together with that of a set of $1800$ surrounding grains in the center of the sample is tracked by a digital video camera triggered in phase with the oscillations of the plate.

\section{Fluidisation and jamming}
\label{sec:Faj}

\subsection{Phase diagram}
\label{sec:Pd}

\begin{figure}[t!] 
\center
\includegraphics[width=0.49\columnwidth,height=0.50\columnwidth]{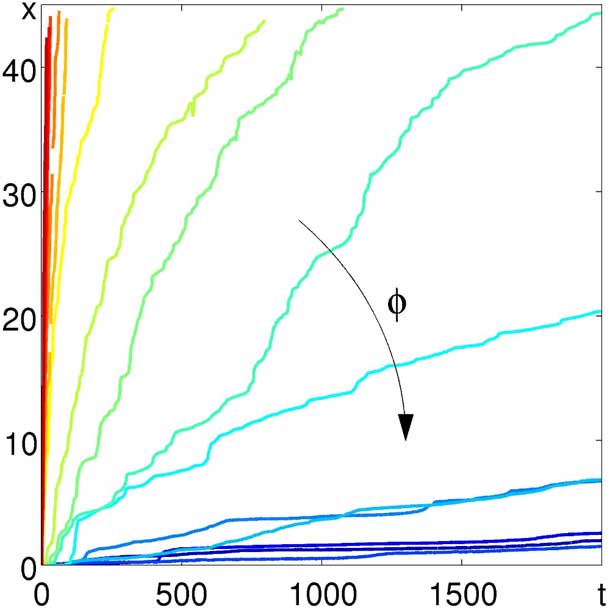}
\includegraphics[width=0.49\columnwidth,height=0.50\columnwidth]{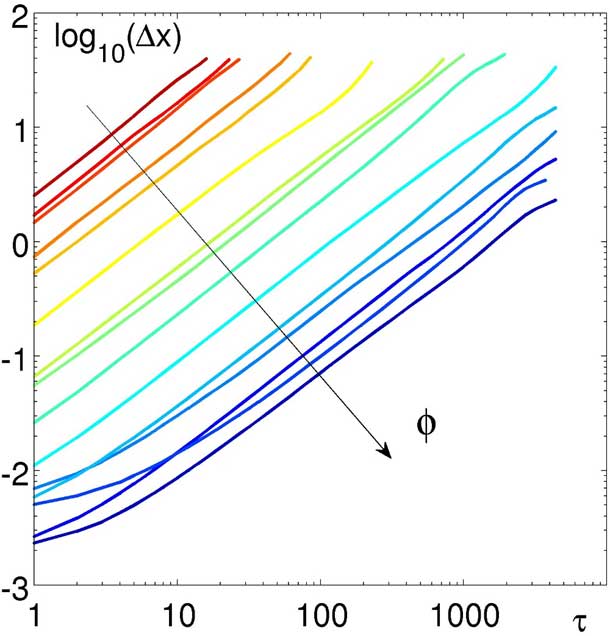}
\includegraphics[width=0.49\columnwidth,height=0.50\columnwidth]{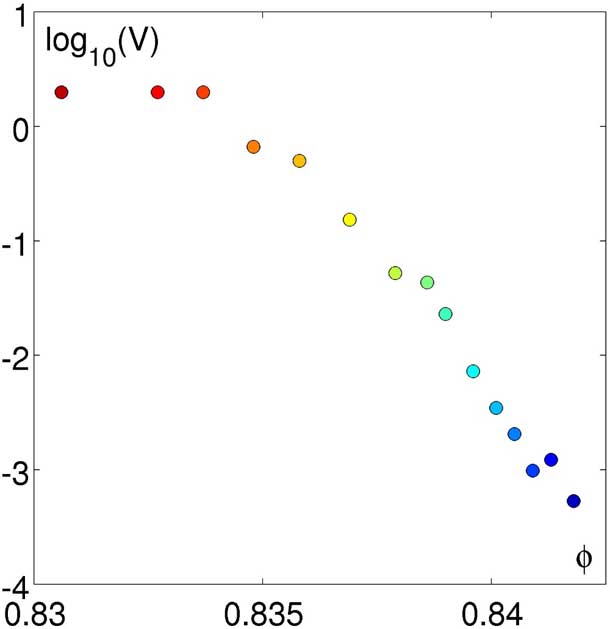}
\includegraphics[width=0.49\columnwidth,height=0.50\columnwidth]{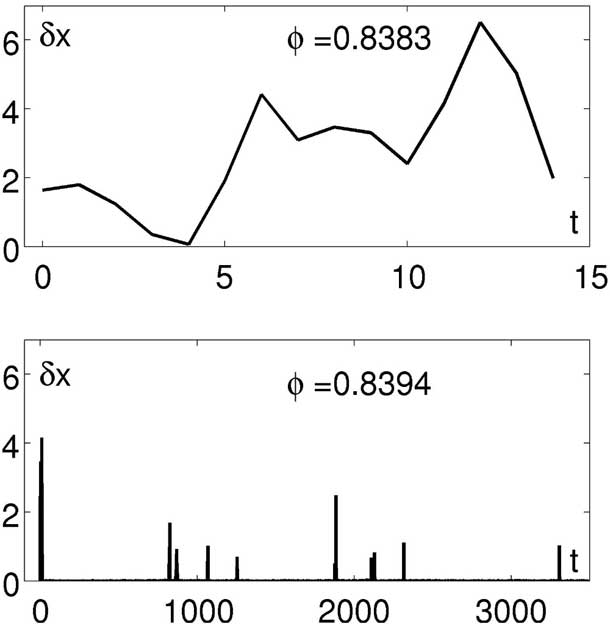}
\caption{
\leg{Top left:} Trajectories of the intruder for several packing fractions, from $\phi=0.8306$ (red) to $\phi=0.8418$ (blue), at constant force ($F=F_2$). The trajectories at the highest packing fractions are truncated.
\leg{Top right:} $\log_{10}$ of the average displacement of the intruder $\Delta x$ in function of the lag time $\tau$ for the same packing fractions and the same drag force.
\leg{Bottom left} $\log_{10}$ of the average velocity of the intruder in function of the packing fraction $\phi$, same packing fractions and drag force.
\leg{Bottom right:} Instantaneous velocities of the intruder at a high drag force ($F=0.363$) for two very close packing fractions on both sides of the fluidisation and jamming transition, here indistinguishable. At $\phi = 0.8383$ the intruder is always moving while at $\phi = 0.8394$ one can observe intermittent bursts of activity separated by long waiting intervals.
}
\label{fig:traj_speed}
\vspace{-0.0cm}
\end{figure}

\begin{figure*}[t!] 
\center
\includegraphics[width=0.66\columnwidth,height=0.66\columnwidth]{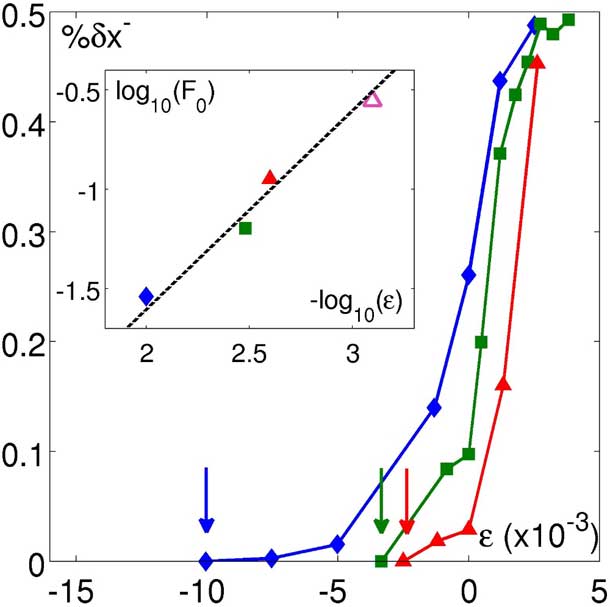}
\includegraphics[width=0.66\columnwidth,height=0.66\columnwidth]{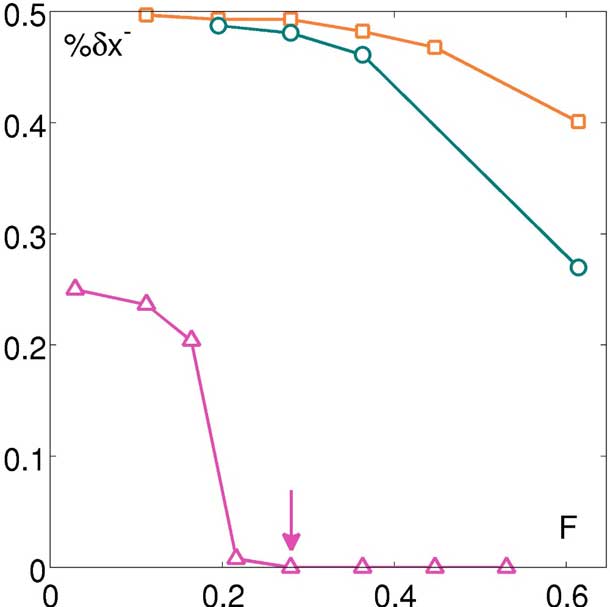}
\includegraphics[width=0.66\columnwidth,height=0.66\columnwidth]{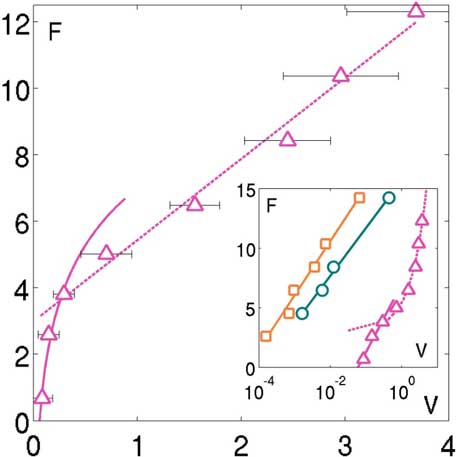}
\caption{
\leg{Left} Ratio of negative displacements $\%\delta x^-$ in function of the reduced packing fraction $\varepsilon$, for three different forces : $F_1$ (\textcolor{bleu}{$\blacklozenge$}), $F_2$ (\textcolor{vert}{$\blacksquare$}) and $F_3$ (\textcolor{rouge}{$\blacktriangle$}). For clarity, only one point with a ratio of $0$ is shown. \inset Fluidisation force in function of the reduced packing fraction $\varepsilon = (\phi-\phi_J)/\phi_J$. Each point correspond to the moment where the ratio of negative displacements reaches $0$. These points are also marked with arrows in the left and middle figures.
\leg{Middle} Ratio of negative displacements in function of the drag force $F$, for three different packing fractions : $\phi_1$ (\textcolor{rose}{$\triangle$}), $\phi_2$ (\textcolor{azur}{$\circ$}) and $\phi_3$ (\textcolor{orange}{$\square$}).
\leg{Right} Applied drag force in function of the average velocity of the intruder in a $lin-lin$ plot at $\phi_1$.
\inset Idem in a $lin-log$ plot for different packing fractions : $\phi_1$ (\textcolor{rose}{$\triangle$}), $\phi_2$ (\textcolor{azur}{$\circ$}) and $\phi_3$ (\textcolor{orange}{$\square$}). Fits are eye-guiding affine (dotted) or logarithmic (plain) behaviors.
}
\label{fig:F_v}
\vspace{-0.0cm}
\end{figure*}

When looking at the trajectories of the intruder along the drag direction $x$ (see fig.\ref{fig:traj_speed}-top left), one immediately notices that the typical velocity dramatically changes within a tiny variation of the packing fraction : the intruder browses the entire system in a few time steps for low packing fractions, and conversely seems to be arrested for the highest values of $\phi$. For a given drag force, the average displacement probed over a lag time $\tau$, $
\Delta x(\tau) = \langle x(t+\tau) - x(t) \rangle_t$ is roughly $V.\tau$, where $V$ is the intruder's average velocity. $V$ spans four decades from $5\times10^{-4}$ to $5$ while varying the packing fraction of only a few percent ($\delta \phi/\phi = 2\times10^{-2}$), illustrating the dramatic freeze of the dynamics (fig.\ref{fig:traj_speed}-top right). For the highest packing fractions, one may notice a systematic bending of the curves at short times indicating that the intruder does not feel instantaneously the bias induced by the drag force.

A closer inspection of the dynamics reveals two salient, distinct regimes of the intruder's motion. For loose packings and large drags, the intruder moves continuously, while for dense packings and small drags the intruder's motion is highly intermittent. Anticipating on the following, let us call this transition ``fluidisation'' and emphasize that it is distinct from the jamming one as illustrated on the phase diagram (fig.\ref{fig:sketch}). Only for the largest forces, both transitions become asymptotically close, as illustrated on figure~\ref{fig:traj_speed}-bottom left, where the instantaneous displacement of the intruder is shown for two packing fractions just below and above the transition. Just below the transition the intruder is moving continuously fast like in a fluid -- note the time-scale on the horizontal axis-- while just above it one observes violent bursts separated by extremely long waiting times, indicating that the medium is not fluidized. We now characterize in more details the nature of these two transitions.

\subsection{Fluidisation}
\label{sec:Fluidisation}

The intruder's motion results from the drag force competing against the resistance of the surrounding grains. When the drag force is low enough -- or if the packing is dense enough -- the configuration can sustain the drag stress until some rearrangement of the force network, induced by the vibration, allows the intruder to move forward. In the meantime, the intruder's motion is cage-like and almost isotropic, going forward and backward roughly half of the time. Figure~\ref{fig:F_v}-left (resp. middle) displays the percentage of time the intruder is going backwards $\%\delta x^-$ as a function of the relative packing fraction $\epsilon$ for the three constant forces $F_1, F_2, F_3$ (resp. as a function of the dragging force for the three packing fractions $\phi_1, \phi_2, \phi_3$). In the extreme cases for which the system is most of the time stuck in jammed states  $\%\delta x^-$ is very close to $0.5$. As the drag force becomes stronger as compared to the resistance of the surrounding grains, the intruder will be less and less often blocked. As a result the percentage of backward steps will be smaller, and eventually will drop to $0$ when the intruder motion becomes continuous. This is precisely how we have chosen to identify the fluidisation transition pointed out by an arrow on the figure. 
The most striking feature is that we could not observe fluidisation for the two packing fractions larger than $\phi_J$. $\%\delta x^-$ decreases with the force, but remains far from zero even for forces as large as $0.6$, that is of the order of the force needed to drag {\it all} the grains on the glass plate in the absence of vibration. Our data suggests a divergence of the fluidisation line at the jamming transition : $F_{flow} \sim \varepsilon^{-1}$ (see inset of figure~\ref{fig:F_v}-left). Looking at the force-velocity relation as shown on figure~\ref{fig:F_v}-right, we observe that in the intermittent motion regime, that is below the fluidisation line, $F\sim \ln(V)$ whereas $F-F_{flow}\propto V$ in the fluidized phase. This demonstrates that the fluidisation line also separates two rheological behaviors. Such a fluidisation transition has been previously reported in other experimental studies. However a straightforward comparison can not be made without further enquiries, which we shall report to the discussion part.

\begin{figure}[t] 
\center
\includegraphics[width=0.49\columnwidth,height=0.48\columnwidth]{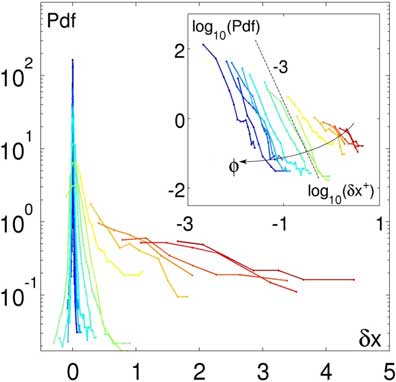}
\includegraphics[width=0.49\columnwidth,height=0.50\columnwidth]{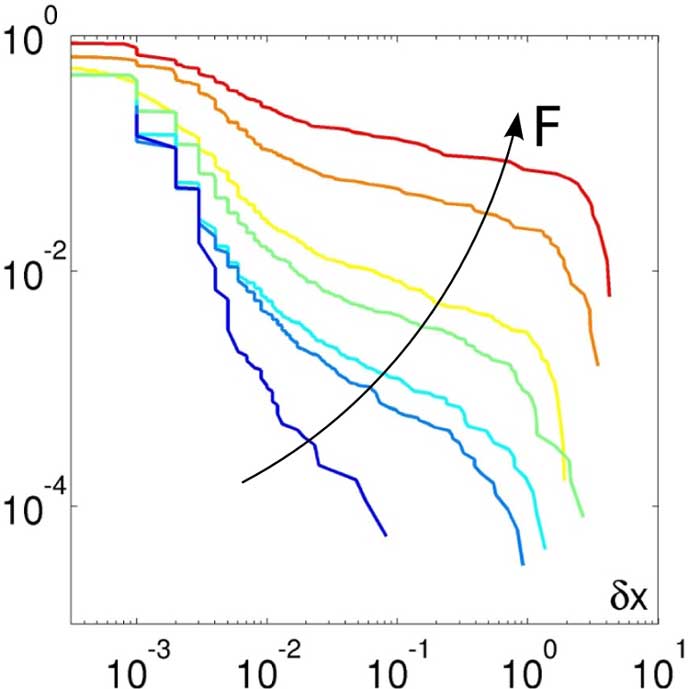}
\includegraphics[width=0.49\columnwidth,height=0.50\columnwidth]{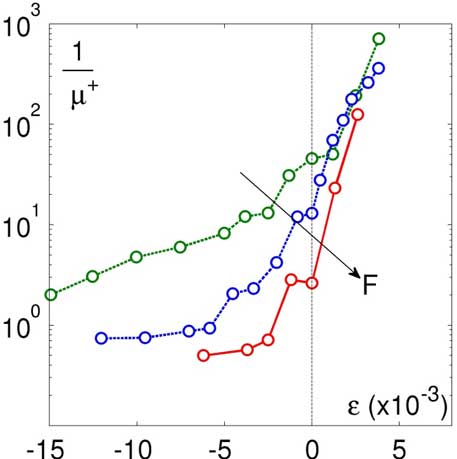}
\includegraphics[width=0.49\columnwidth,height=0.48\columnwidth]{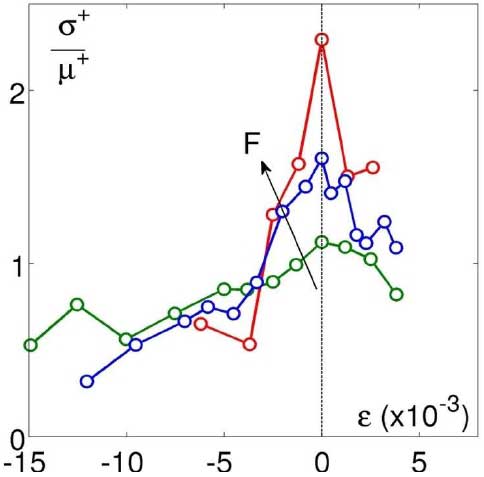}
\caption{
\leg{Top left} Pdf of the instantaneous displacements in the drag direction $\delta x$ at a constant force $F=F_2$ for several values of $\phi$.
\inset same for the positive displacements only, in $log-log$ scale.
\leg{Top right} Cumulated Pdf of the instantaneous displacements in the drag direction $\delta x$ at a constant packing fraction $\phi=\phi_1$ for several values of $F$ (right).
\leg{Bottom} Inverse average $1/\mu^+$ (left) and standard deviation over average $\sigma^+/\mu^+$ (right) of the instantaneous positive displacements in the drag direction as functions of the reduced packing fraction $\varepsilon$. Different curves correspond to different drag forces : $F_1$ (green), $F_2$ (blue) and $F_3$ (red).}
\label{fig:PDF_fluc}
\vspace{-0.0cm}
\end{figure}

\subsection{Jamming}
\label{sec:Jamming}

We now characterize the intruder dynamics when approaching the jamming transition. When looking at the intruder's instantaneous displacements $\delta x$ along the drag direction in the intermittent regime (fig.\ref{fig:traj_speed}-bottom right and fig.\ref{fig:crackling_sketch}-left), one immediately notices very strong fluctuations, with bursts of widely fluctuating magnitude in the direction of the drag. As a result the distributions of $\delta x$ (fig.\ref{fig:PDF_fluc}-top left) exhibit an important skewness toward the positive displacements for the packing fractions above fluidisation. In order to characterize these positive displacements $\delta_x^+$, we compute the average value over time $\mu^+=\langle\delta x^+\rangle_t$ and the relative fluctuations~:
\be
\frac{\sigma^+}{\mu^+} = \frac{\sqrt{\langle(\delta x^+ - \mu^+)^2\rangle_t}}{\mu^+}.
\ee
\noindent
Figure~\ref{fig:PDF_fluc}-bottom left shows that $1/\mu^+$, the typical time the intruder takes to move one particle diameter, increases monotonically with $\varepsilon$ and faster than exponentially. The stronger the dragging force the sharper the increase. No significant behavior is observed when crossing the jamming transition. On the contrary, figure~\ref{fig:PDF_fluc}-bottom right reveals a peak of $\sigma^+/\mu^+$ precisely at $\phi_J$ for the three dragging forces. 
Note that the peak sharpens when the drag force is stronger. Indeed, as discussed in the previous section, when $F$ increases the fluidisation line approaches the jamming transition. This has two consequences. The displacement bursts become larger when they occur -- as can be seen on figure~\ref{fig:PDF_fluc}-top left from the cumulated distribution of the intruder displacements for increasing forces -- and the range of packing fraction separating the non fluctuating continuous motion in the fluidized regime from the strongly intermittent one at jamming shrinks. Finally let us mention 
that in the intermittent regime the distributions of $\delta x^+$ decay as power-laws with an exponent close to $-3$ (see inset of figure~\ref{fig:PDF_fluc}-top left), a result similar to the one reported in a recent simulation, where a probe is dragged into an assembly of harmonically repulsing disks~\cite{reichhardt2009fjy}. 

The above results reveal that the strong spatio-temporal heterogeneities of the dynamics reported in the absence of intruder~\cite{lechenault2008csa} can also be seen in the response of the intruder to the dragging force. This is related to the fact that microrheology gives access to non-linear responses and thereby probes the strength required to pull free the probe from the transient
local structure. Before addressing in further details the critical nature of the fluctuations reported here above, we will concentrate on the response of the grains surrounding the intruder. This will ultimately allow us to perform the statistical analysis of the rearrangement events as a whole, involving both the intruder and the surrounding grains.

\section{Around the intruder}
\label{sec:Ati}

\subsection{Average flow and spatial fluctuations}
\label{sec:Afasf}

\begin{figure}[t!] 
\centering
\includegraphics[width=0.49\columnwidth]{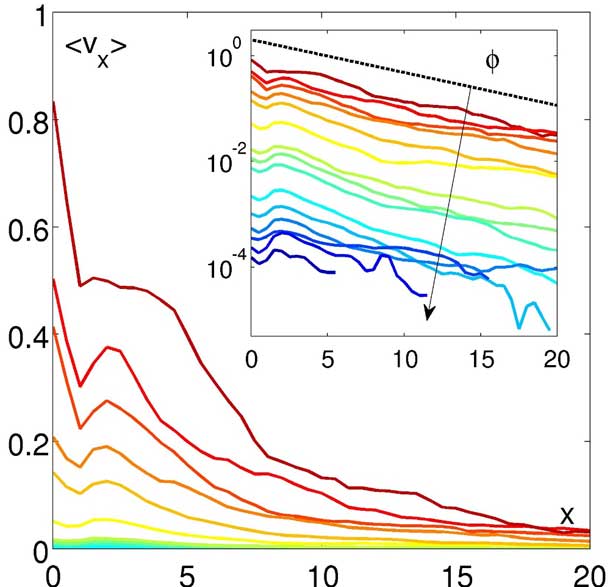}
\includegraphics[width=0.49\columnwidth]{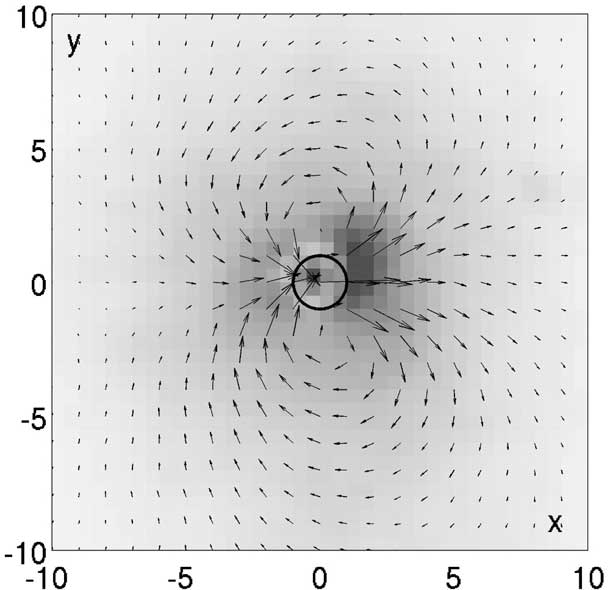}
\includegraphics[width=0.49\columnwidth]{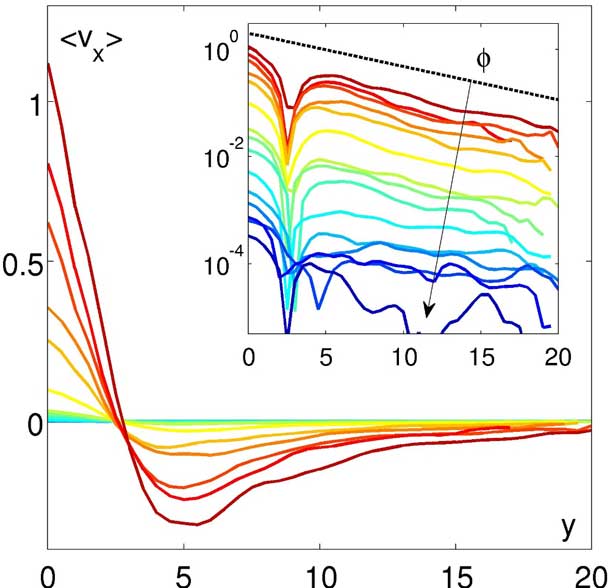}
\includegraphics[width=0.49\columnwidth]{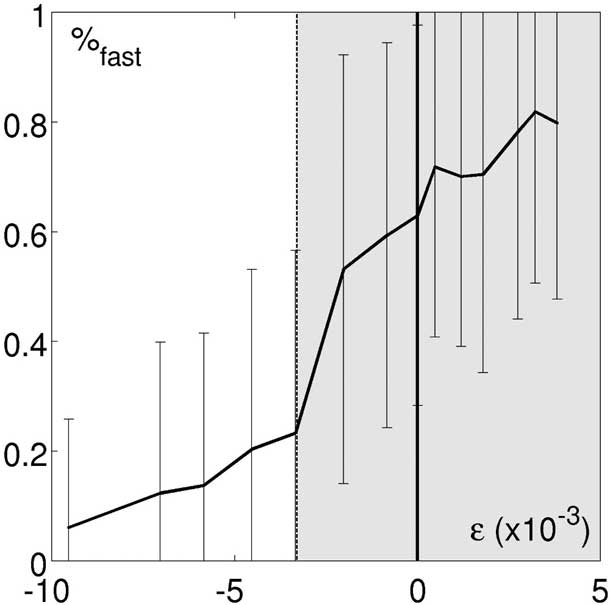}
\caption{
\leg{Left} Velocity profiles : average velocity in the $x$-direction along the horizontal (top) and vertical (bottom) axis, for $15$ packing fractions.
\inset Same in absolute values and $log-lin$. The dark dotted line is an exponential decaying guide with a characteristic length $\lambda_x = 7$.
\leg{Top-right} Displacement field around the intruder at $\phi=0.8386$, averaged over time. The arrows are magnified by a factor 3.
\leg{Bottom-right} Average percentage of particles in the connex clusters of particles faster than $v_{max}/2$ in function of the reduced packing fraction. Error bars show the standard deviation. The grey area correspond to the intermittent regime. In all plots, $F=F_2$.
}
\label{fig:disp_fields}
\vspace{-0.0cm}
\end{figure}

As already suggested in figure~\ref{fig:setup}, the instantaneous displacement field around the intruder during a burst is rather complex. They are typically asymmetric, with a main vortex on one or the other side of the intruder. Averaged over many bursts (see figure~\ref{fig:disp_fields}-top right), the displacement field recovers the intruder's left-right symmetry and exhibit a two vortices pattern. 
Further insight into the dependence of this flow on the packing fraction is obtained when looking at the profiles of the $x$-component of the velocity along the direction of the drag ($x$-direction) and perpendicular to the drag ($y$-direction) (see figure~\ref{fig:disp_fields}-right top and bottom).
As evidenced in the insets, one observes that the velocity field decreases exponentially with the distance to the intruder, and amazingly that the typical length-scales $\lambda_x \simeq 7$ and $\lambda_y \simeq 10$ associated with such a dependence are totally independent from the packing fraction. This is also confirmed by the localization of the center of the vortices, easily located on the $y$-profile, which remains at the same distance from the intruder at all packing fractions.
Hence, as far as the averaged flow is concerned, only its overall magnitude depends on the packing fraction and scales like $V$, the averaged velocity of the intruder.

In order to study the fluctuations of the displacement field, one could imagine to subtract the averaged field; however without a prescription for the analytical form of this field or better statistics, this is a rather uncontrolled operation. An alternative procedure is to define first a maximal displacement for the grains, $v_{max}$, as the average displacement of the intruder's closest neighbors, and then to define the ``fast particles'' cluster as the connex cluster of particles around the intruder which move more than $v_{max}/2$. The average number of particles in this fast cluster is plotted on figure~\ref{fig:disp_fields}-bottom right as a function of the packing fraction: it grows from $10\%$ to $80\%$ in the tiny interval between $\phi=0.8306$ and $\phi=0.8409$, underlining how sharply the spatial extent of the reorganizations grows as the system goes toward dynamical arrest.
It is however surprising that the number of fast particles increases, considering that the averaged flow scales entirely with the velocity of the intruder which we have scaled out by defining the ``fast'' particles relatively to the average velocity around the intruder. Such a difference between the averaged flow behavior and the instantaneous one can only be explained by the existence of strong heterogeneities in the instantaneous fields. The amplitude of the fluctuations of the number of fast particles, as indicated by the error bars on figure~\ref{fig:disp_fields}-bottom right is already an indication that it is indeed the case.

\begin{figure}[t]
\center
\includegraphics[width=0.32\columnwidth]{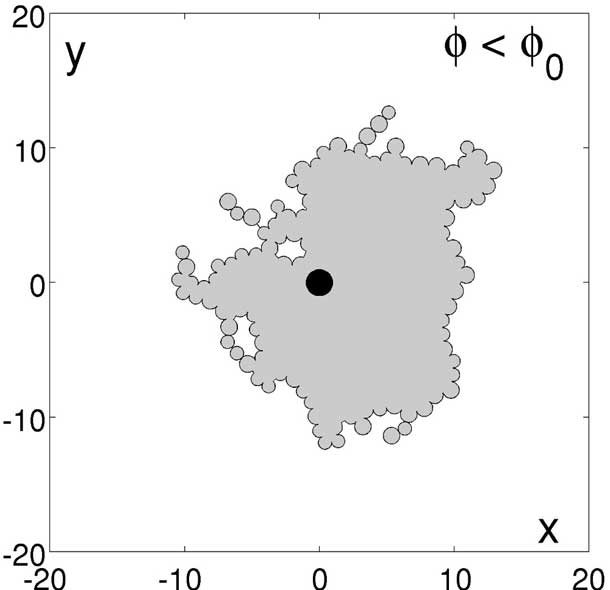}
\includegraphics[width=0.32\columnwidth]{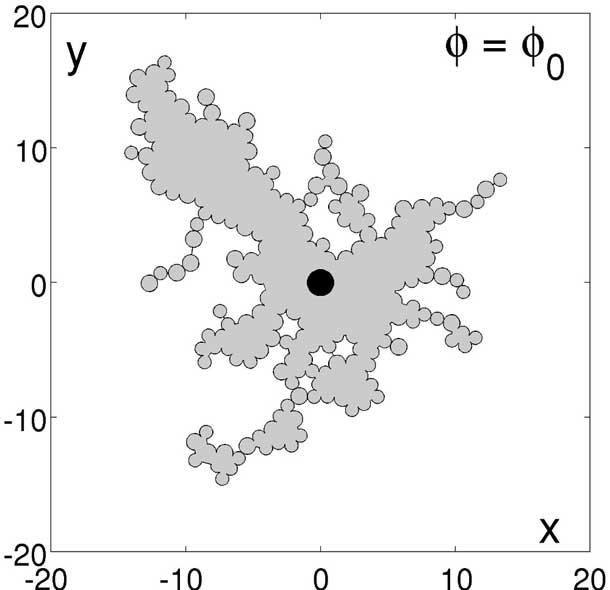}
\includegraphics[width=0.32\columnwidth]{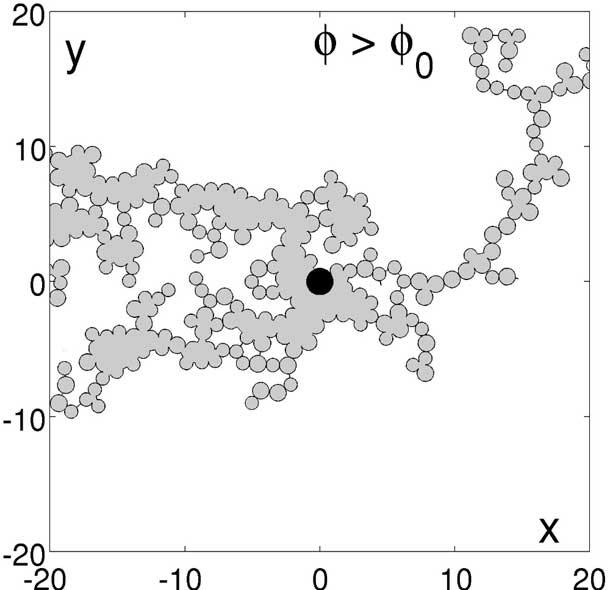}
\includegraphics[width=0.48\columnwidth]{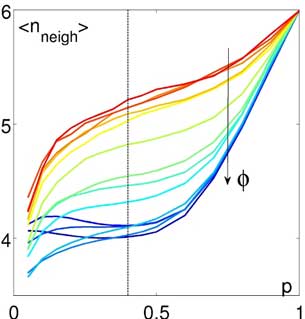}
\includegraphics[width=0.48\columnwidth]{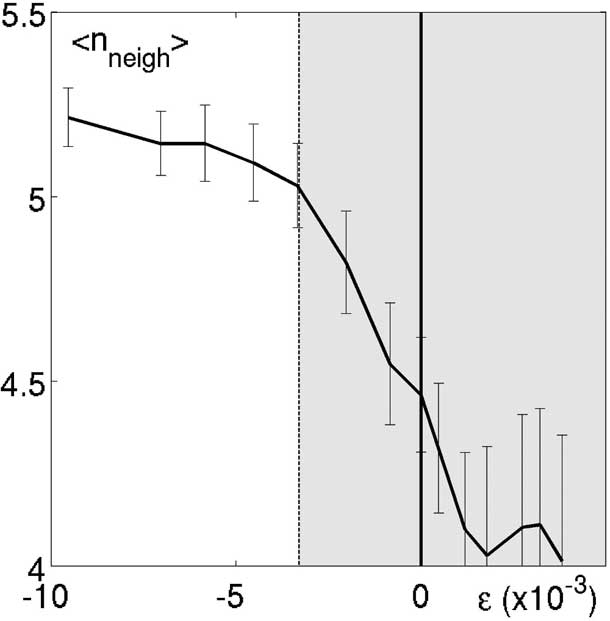}
\includegraphics[width=0.48\columnwidth]{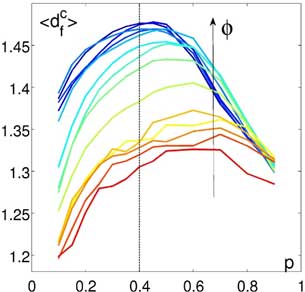}
\includegraphics[width=0.48\columnwidth]{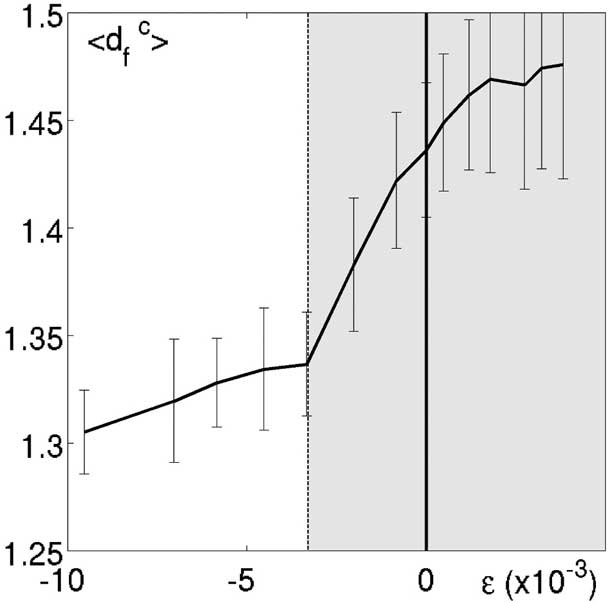}
\caption{
\leg{Top} Typical contours of clusters of the $p=15\%$ fastest particles, for three packing fractions, from left to right~: $\phi=0.8337$ ($\phi < \phi_0$), $\phi=0.8358$ ($\phi = \phi_0$) and $\phi=0.8396$ ($\phi > \phi_0$). The intruder is marked by a black disk and moves from left to right.
\leg{Middle left} Average number of neighbors  $\langle n_{neigh} \rangle$ in function of $p$, for different packing fractions. The black dotted line is at $p=40\%$, in the middle of the plateau.
\leg{Middle right} $\langle n_{neigh} \rangle$ in the clusters of $p=40\%$ fastest particles, as a function of $\varepsilon$. The grey area correspond to the intermittent regime, and error bars show the standard deviation.
\leg{Bottom left} Average fractal dimension of the clusters contours $\langle d_f^c \rangle$ as a function of $p$ for different packing fractions. The black dotted line is at $p=40\%$.
\leg{Bottom right} $\langle d_f^c \rangle$ as a function of $\varepsilon$ for $p=40\%$. The grey area correspond to the intermittent regime, and error bars show the standard deviation.
$F=F_2$ for all plots.
}
\label{fig:shape_df}
\vspace{-0.0cm}
\end{figure}

A further characterization of these heterogeneities is provided by the shape of the clusters made of the $p\%$ fastest particles at each time step; typical examples with $p=15\%$ are shown on fig.\ref{fig:shape_df}-top for three packing fractions below, at, and above $\phi_0$ the fluidisation packing fraction. One clearly observes that these instantaneous rearrangements evolve with the packing fraction from dense to branched patterns with long chains spanning the whole system.
The number of neighbors $n_{neigh}$ inside these clusters is a good indicator of the level of branching since its typical value is $6$ for perfectly dense clusters and $2$ for perfect strings. 
The well defined plateau in the dependence of $\langle n_{neigh}\rangle$ with $p$ (figure~\ref{fig:shape_df}-middle center) corresponds to the clusters which contain enough particles to have a non-trivial shape but do not reach the boundaries of the acquisition field. The average number of neighbors at the plateau (fig.\ref{fig:shape_df}-middle right) decreases significantly with the packing fraction providing a quantitative evaluation of the clusters' evolution from dense aggregates to branched structures. The contour fractal dimension $d_f^c$ calculated with the classical compass method~\cite{kaye1989arw} also gives insight on branching. The mean value $\langle d_f^c\rangle(p)$ has a nice maximum for the range of $p$ corresponding to the plateau in $\langle n_{neigh}\rangle(p)$. For a value of $p=40\%$, that is for clusters large enough to assess the fractal dimension without spanning the acquisition field, we observe a significant increase of $d_f^c(\phi)$ at $\phi_0$, the packing fraction where the intruder's motion becomes intermittent.

Altogether, the patterns observed in the instantaneous displacement field during the relaxation events suggest that the force chains network, whose importance has been visually exemplified in photo-elastic disks experiments of an intruder dragged in a Couette cell~\cite{majmudar2007tjt}, starts playing a significant role as soon as the system crosses the fluidisation transition. However, one can not elude the role of the density field, which we now investigate computing the free volume field around the intruder.

\subsection{Free volume}
\label{sec:Fv}

\begin{figure}[t] 
\center
\includegraphics[width=0.49\columnwidth]{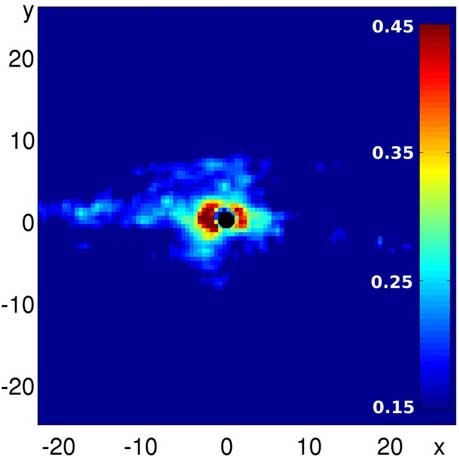}
\includegraphics[width=0.49\columnwidth]{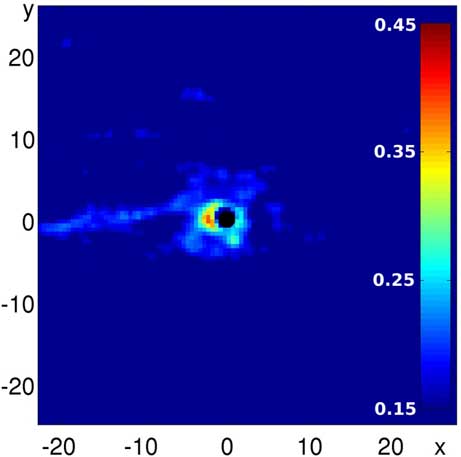}
\includegraphics[width=0.49\columnwidth]{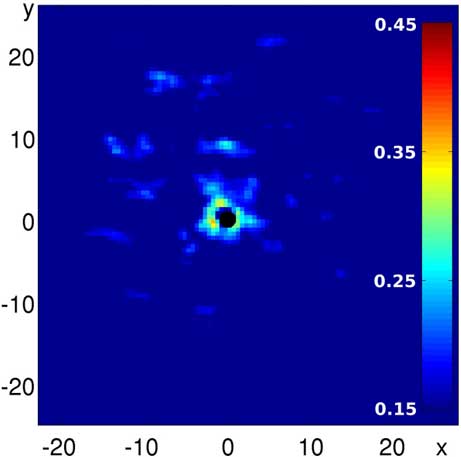}
\includegraphics[width=0.48\columnwidth]{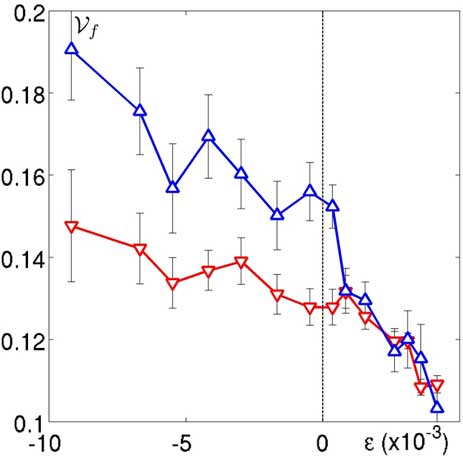}
\caption{
\leg{Top and bottom-left} Average free volume fields around the intruder for three packing fractions: $\phi=0.8327$ ($\phi<\phi_J$, top-left), $\phi=0.8386$ ($\phi\simeq\phi_J$, top-right), $\phi=0.8405$ ($\phi>\phi_J$, bottom-left).
\leg{Bottom-right} Average free volume $\langle \mathcal{V}_f \rangle$ before (\textcolor{rouge}{$\triangledown$}) and after (\textcolor{bleu}{$\vartriangle$}) the intruder in function of the reduced packing fraction. Error bars represent the standard deviation over time of the free volume in each point of space, averaged on the computation window.
}
\label{fig:FV}
\vspace{-0.0cm}
\end{figure}

We then examine the density field around the intruder. For that purpose, we compute the free volume field extracted from Laguerre's tesselation~\footnote{Laguerre's tesselation is similar to Delaunay's, but power distances are used instead of distances in order to take polydispersity into account.} of the configurations at each time step: the free volume $\mathcal{V}_f$ of a Vorono\"i cell is the area difference with the minimal possible regular hexagon around the grain normalized by the surface of the grain. Figure~\ref{fig:FV} displays the averaged free volume field around the intruder for three packing fractions, below, above, and at the jamming transition. Above $\phi_J$, there is a very small amount of decompaction around the intruder without significant signature of the intruder motion, seemingly an effect of the size difference between the intruder and the surrounding grains disturbing the local organization of the packing. On the contrary, as the packing fraction decreases below jamming, two holes grow, first on the back of the intruder and then on its front.

As a consequence, one observes a clear signature of the intruder's motion, namely the apparition of a forward-backward asymmetry which can be quantified by computing for instance the average free volume in front of and behind the intruder (figure~\ref{fig:FV}-bottom right)\footnote{The averaging windows are rectangles that start at the position of the intruder, with a height of $10$ diameters of small grains and the maximal possible length}. One observes that above $\phi_J$, there is no more sign of the asymmetry associated with the intruder motion.

Altogether, we have seen that the averaged flow around the intruder is surprisingly insensitive to the transition, apart from a simple scaling factor indexed on the average velocity of the probe. On the contrary, the fluctuations reveal a rather complex interplay of the density and stress fluctuations. Below the fluidisation transition, the intruder motion is dominated by important free volume rearrangements, which concentrate in dense clusters of fast moving grains around the intruder. As soon as the system enters the intermittent regime, these clusters start to branch, indicating the existence of inhomogeneities in the rigidity of the system. The instantaneous displacements spread on larger and larger scales; and progressively the free volume fluctuations become distributed throughout the system. Above $\phi_J$ these last type of rearrangement completely dominate the intruder motion.

\section{Burst statistics close to jamming}
\label{sec:Bsctj}

\begin{figure}[t!] 
\center
\includegraphics[width=0.49\columnwidth]{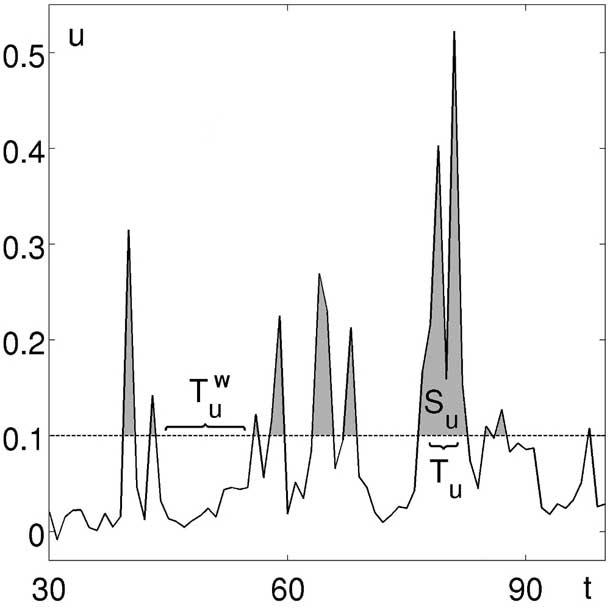}
\includegraphics[width=0.49\columnwidth]{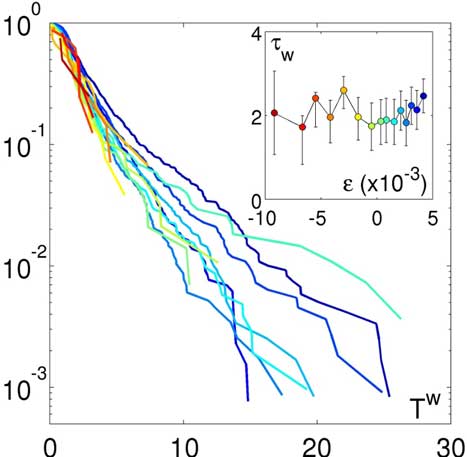}
\includegraphics[width=0.49\columnwidth]{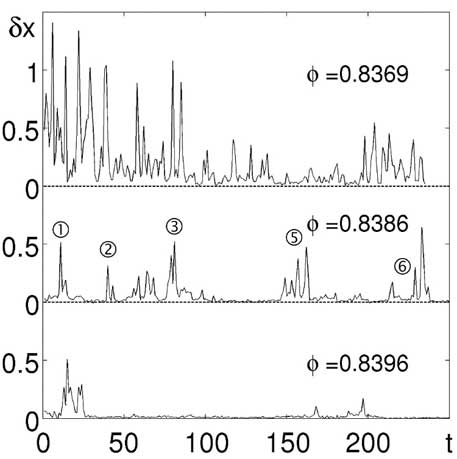}
\includegraphics[width=0.49\columnwidth]{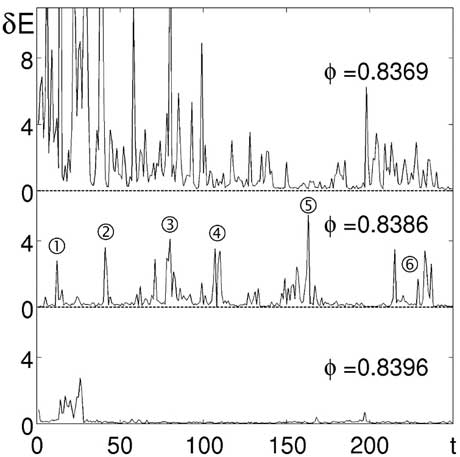}
\caption{
\leg{Top left} Definition of the bursts of activity : given a threshold on a signal $u$, each connex portion of signal above the threshold defines a burst. $T_u$ is the duration of the burst while $S_u$ is its area (in grey). The waiting times between bursts are noted $T_u^w$.
\leg{Top Right} Cumulated probability density functions of the time intervals $T^w$ between bursts of displacements for $15$ packing fractions from  $\phi=0.8306$ (red) to $0.8418$ (blue), at the constant drag force $F_2$. \textit{Inset} Average waiting time $\tau_w$ for the same packing fractions estimated by an exponential fit. Error bars represent the $95\%$ confidence bounds of the fit.
\leg{Bottom left} Instantaneous displacements $\delta x(t)$ of the probe for three packing fractions.
\textbf{Bottom right} Energy of the grains $\delta E(t)$ around the intruder for the same packing fractions. Note that all peaks can be found in both signals, except some rare events like the peak number $4$. $F=F_2$ for all bottom plots.
}
\label{fig:crackling_sketch}
\vspace{-0.0cm}
\end{figure}

In the above section, we have seen that the temporally intermittent motion of the intruder also corresponds to spatially heterogeneous rearranging regions. We will now investigate the intruder displacements and the kinetic energy of these regions, focusing on the temporal correlations at the root of this dynamics of bursts. Figure~\ref{fig:crackling_sketch}-bottom displays the instantaneous displacement $\delta x(t)$, and the kinetic energy involved in a rearrangement $\delta E(t) = \Sigma \delta x_i^2(t)$, where the sum is performed on the clusters of particles faster than $v_{max}/2$, as defined in section \ref{sec:Afasf}. For three different packing fractions chosen above the fluidisation transition, one clearly observes a sequence of distinguishable and well separated bursts, suggesting ``crackling noise''-like signals~\cite{sethna2001cn, durin2006tbe}.

Note that the signals of $\delta x(t)$ and $\delta E(t)$ are very similar. However the correlation is not as strong as one would first imagine by visual inspection. Each burst in the displacements signal corresponds indeed to a burst in the kinetic energy one but the bursts might be shifted in time and the intruder's displacements sometimes precede, sometimes follow the energy ones. In addition, the amplitude of the bursts only weakly correlate and there are some occasions where there is a burst of activity in the system without a displacement of the intruder itself (see an example on fig.~\ref{fig:crackling_sketch}), but these are marginal events.

The first step in analyzing such kind of signals is to coarsegrain the statistical distribution of the jumps to capture the size and duration of the relevant bursts. One common way to do so is to define a threshold, or resolution coefficient, in order to delimit the temporal limits of each burst : given a signal $u(t)$ and a threshold $\hat{u}$, one can define bursts or avalanches as the connex portions of the signal that stand above $\hat{u}$~; every burst have a duration $T_u^i$ and an integrated amplitude $S_u^i$ (see fig.\ref{fig:crackling_sketch}-top left).
Choosing the threshold is a delicate problem given the large variations of the displacements' amplitudes, the complete loss of small events for the loosest packings and the fact that the background noise limits the detection of small bursts at high packing fractions. In the following we have set all the thresholds at the value where the difference between the cumulated distributions of the local minimums and maximums of $u(t)$ exhibits a peak. On one hand this corresponds to the point where the minimums and maximums in $u(t)$ are best separated, and on the other hand it points out to the threshold value for which the number of bursts is maximal, naturally enforcing the statistics. We have checked that multiplying the so-obtained thresholds by a factor from $1/2$ to $2$ doesn't change the following results.

\begin{figure*}[t!] 
\centering
\includegraphics[width=0.65\columnwidth]{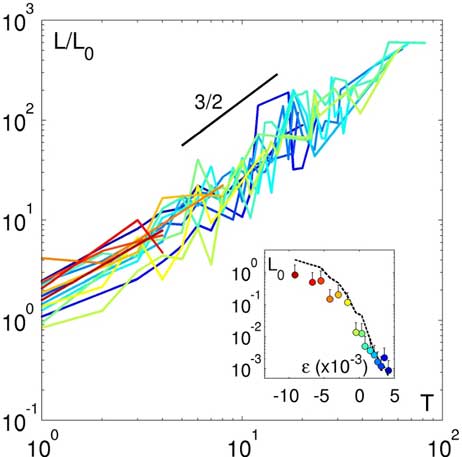}
\includegraphics[width=0.65\columnwidth]{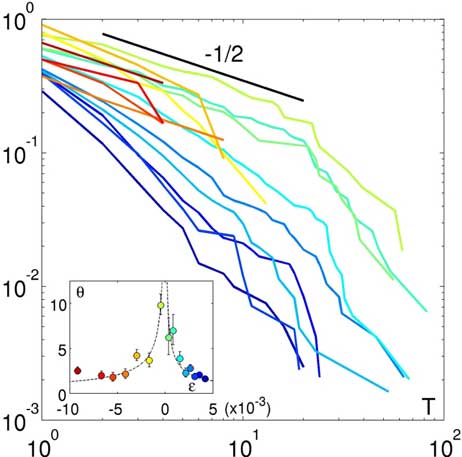}
\includegraphics[width=0.65\columnwidth]{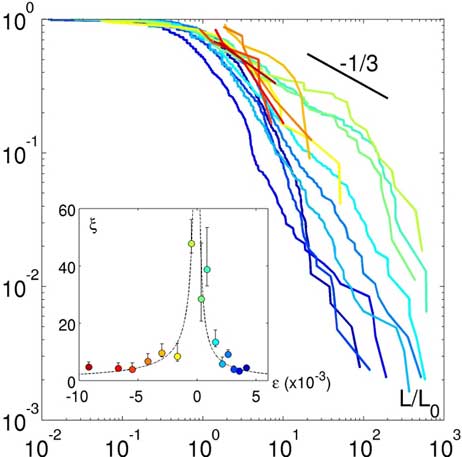}
\includegraphics[width=0.65\columnwidth]{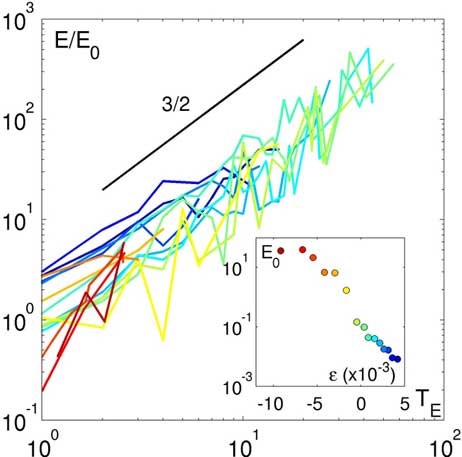}
\includegraphics[width=0.64\columnwidth]{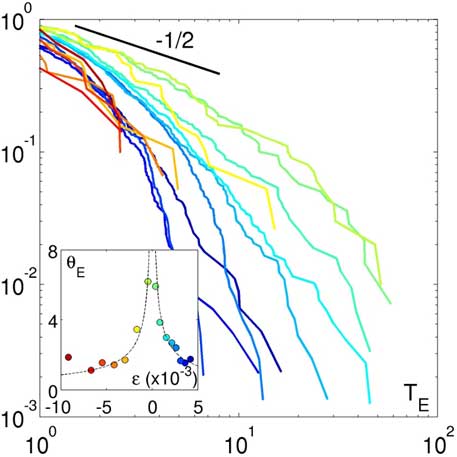}
\includegraphics[width=0.65\columnwidth]{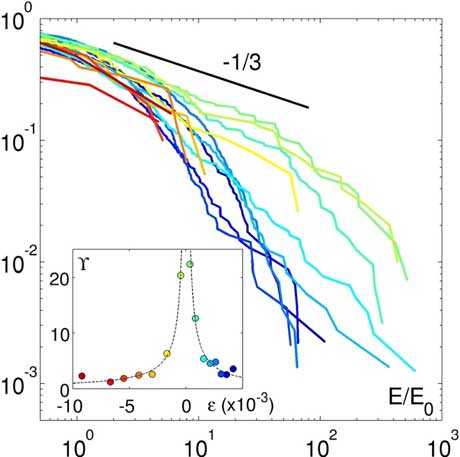}
\caption{
\leg{Top left:} Rescaled length $L/L_0$ as a function of $T$. \textit{Inset} Scaling factor $L_0$ as a function of $\varepsilon$. The black dotted line corresponds to the average speed of the intruder.
\leg{Top middle:} Cumulated Pdf of the burst durations $T$ in the intruder's displacement signal. \textit{Inset} Cut-off $\theta(\varepsilon)$ of the distribution of $T$. The eye-guiding lines are power laws with an exponent $-2/3$.
\leg{Top right:} Cumulated Pdf of the rescaled length $L/L_0$. \textit{Inset} Cut-off $\xi(\varepsilon)$ of the distribution of $L/L_0$. The eye-guiding lines are power laws, with an exponent $-1$.
\leg{Bottom left:} Rescaled energy $E/E_0$ as a function of $T_E$. \textit{Inset} Scaling factor $E_0$ as a function of $\varepsilon$.
\leg{Bottom middle:} Cumulated Pdf of the burst durations $T_E$ in the energy signal. \textit{Inset} Cut-off $\theta_E(\varepsilon)$ of the distribution of $T_E$. The eye-guiding lines are power laws with an exponent $-2/3$.
\leg{Bottom right:} Cumulated Pdf of the rescaled energy $E/E_0$. \textit{Inset} Cut-off $\Upsilon(\varepsilon)$ of the distribution of $E/E_0$. The eye-guiding lines are power laws, with an exponent $-1$.
}
\label{fig:avalanches}
\vspace{-0.0cm}
\end{figure*}

First we observe on figure~\ref{fig:avalanches}-top right that the waiting times separating two succesive bursts are exponentially distributed the characteristic time of which $\tau_w \simeq 2$ is independent of the packing fraction (see inset). Two conclusions can be drawn: $(i)$ the bursts follow a Poissonian process and can be considered as independent events, and $(ii)$ $\tau_w$ remains very small compared to the total time of the experiments. We catch a large number of bursts, even at the highest packing fractions. Note that the use of a constant threshold to define the bursts would have lead to a dramatic increase of these waiting times in the intermittent regime and for the highest packing fractions the number of bursts would have been vanishingly small.

We focus now on the statistics of the bursts themselves. Let us note $T$ and $L$, respectively $T_E$ and $E$, the duration and the integrated amplitude of the bursts recorded on the signal of the displacements of the probe, respectively of the kinetic energy of the surrounding fastest grains. Obviously these quantities are not independent one from another. The first step in the analysis (see figure~\ref{fig:avalanches}-left), shows that :
\begin{eqnarray}
L(T,\phi)= L_0(\phi) T^{1/z}\quad \textrm{and}\quad E(T_E,\phi)= E_0(\phi) T_E^{1/z},
\label{eq:TLz}
\end{eqnarray}
where $L_0$ and $E_0$ can be interpreted as the typical displacement and energy associated with a burst and $z=2/3$ is usually called the \textit{dynamical exponent}. $L_0$ depends on $\phi$ in a similar way as the average velocity of the intruder does, consistently with the previous observation that the average waiting time $\tau_w$ is independent from $\phi$.

The second observation concerns the distributions of $T, T_E$ and $L/L_0, E/E_0$. Again we find identical behaviors for the intruder displacement and the kinetic energy of the fastest surrounding grains (fig.~\ref{fig:avalanches}-middle and right). All the above quantities are largely distributed, with a large value cutoff, which depends on the packing fraction. Visual inspection of the tails of the distribution already indicates that the cutoff dependence on the packing fraction is not monotonic. In all plots the jamming packing fraction corresponds to the light green curves. The distributions are the largest for that precise packing fraction. Such behaviors can be encoded in the following scaling relations:
\begin{eqnarray}
\rho(T) & \propto & T^{-(1+\alpha)}.f \left( \frac{T}{\theta(\varepsilon)} \right),
\label{eq:rho_T}\\
\rho(T_E) & \propto & T_E^{-(1+\alpha)}.f_E \left( \frac{T_E}{\theta_E(\varepsilon)} \right), 
\label{eq:rho_TE}\\
\rho(L/L_0) & \propto & (L/L_0)^{-(1+\beta)}.g \left( \frac{L}{\xi(\varepsilon)} \right),
\label{eq:rho_L}\\
\rho(E/E_0) & \propto & (E/E_0)^{-(1+\beta)}.g_E \left( \frac{E}{\Upsilon(\varepsilon)} \right), 
\label{eq:rho_E}
\end{eqnarray}
where the exponents $\alpha \simeq 1/2$ and $\beta \simeq 1/3$ satisfy the expected relation $\alpha.z = \beta$. Note that given the lack of statistics these exponents are not the results of a fit but only indicative values.  One then easily extracts the scaling variables, which are the cutoffs of the distributions and are simply proportional to their mean values~\footnote{Given a random variable $x$, if the distribution of $x$ is a power law with a cut-off $b$, \ie of the form $\rho(x) = x^{-1+a}.f(x/b)$, the average value of $x$ is~:
$\langle x \rangle = b \times \frac{\int_{0}^{+\infty} u^{-a}f(u).du}{\int_{0}^{+\infty} u^{-(1+a)}f(u).du} = b \times cst$. One can therefore reproduce the evolution of the cut-offs as a function of the packing fraction by taking the experimental averages.}. Their dependence on the packing fraction are plotted in the insets of figure~\ref{fig:avalanches} and exhibit a sharp peak at the jamming transition, suggesting a critical behavior and thereby ensuring the self-consistency of the above scaling analysis. This critical behavior is well described by:
\begin{eqnarray}
\theta(\varepsilon) \propto \theta_E(\varepsilon) \propto \varepsilon^{-\eta},
\label{eq:theta_eta}\\
\xi(\varepsilon) \propto \Upsilon(\varepsilon) \propto \varepsilon^{-\nu},
\label{eq:xi_nu}
\end{eqnarray}
where $\eta \simeq -2/3$ and $\nu \simeq -1$ also satisfy the relation $\eta = \nu.z$.

\begin{figure}[!t] 
\center
\includegraphics[width=0.75\columnwidth]{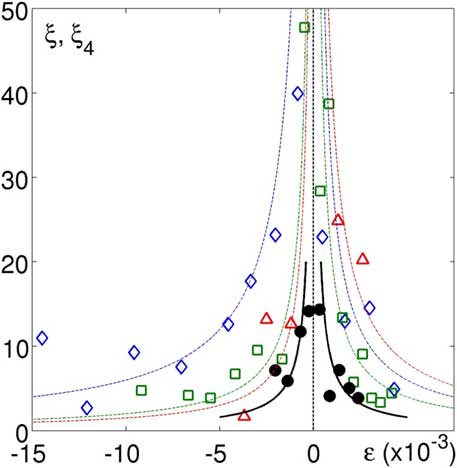}
\caption{
Comparison of the evolution of dynamical lengthscales~: the cut-off of the distribution of the bursts of displacement's amplitudes $\xi$ is plotted for $3$ different forces (\textcolor{bleu}{$\Diamond$} $F_1$, \textcolor{vert}{$\Box$} $F_2$ and \textcolor{rouge}{$\triangle$} $F_3$) and compared to ($\bullet$) the long-range correlation length $\xi_4$, revealed from the dynamics of the grains without the intruder. The dotted and plain curves are guiding the eye like $\varepsilon^{-1}$.
}
\label{fig:criticality}
\vspace{-0.0cm}
\end{figure}

Altogether the above analysis provides strong evidences of a critical behavior of the intruder motion at the jamming transition, enforcing the first evidences of dynamical criticality reported in \cite{lechenault2008csa}. Figure~\ref{fig:criticality} reports the lengthscale $\xi$ measured here for the three different dragging forces, together with the dynamical correlation length $\xi_4$, which is the correlation length of the local density relaxations, and which was computed in \cite{lechenault2008csa}. The authors reported a dependence of $\xi_4$ with the distance to jamming compatible with $\xi_4 \sim \varepsilon^{-1/2}$, however it can not really be discriminated from $\xi_4 \sim \varepsilon^{-1}$, as reported in the present work for the lengthscale $\xi$. 
The fact that both lengthscales behave in a similar way at the transition suggests a fluctuation-dissipation-like relation between the non-linear response studied here and dynamical heterogeneities, as recently discussed in the context of Mode Coupling Theory~\cite{tarzia2008anl}.
In the following we discuss our results as compared to other microrheological studies in dense systems of particles, and try to precise the kind of criticality we are facing at jamming.

\section{Discussion and concluding remarks}
\label{sec:Dacr}

We have experimentally studied the motion of an intruder dragged into an amorphous monolayer of horizontally vibrated grains at high packing fractions. A first ``fluidisation'' transition separates a continuous motion regime, where the force-velocity relation is affine, from an intermittent motion one, where the force depends logarithmically on the velocity. The force threshold increases with the packing fraction and seemingly diverges at the jamming transition defined as the packing fraction where the pressure goes to zero in the absence of vibration. Below this threshold the intruder motion is intermittent. We have reported the existence of a second transition, where the fluctuations associated with this intermittent motion exhibit a critical behavior. This evidence is supported by the analysis of the motion of the probe but also by that of the surrounding grains. The bursting events are characterized by increasingly heterogeneous patterns in the instantaneous displacement field around the intruder and an increasingly spreading redistribution of the free volume. 

Several other experimental, numerical and theoretical investigations of the response of a locally driven particle in a dense system report similar observations. Our aim here is to discuss similarities and differences among these studies, in an attempt to shed some light on sometime apparently controversial results. To our knowledge the first measurements were conducted in colloidal suspensions~\cite{habdas2004fmp}. At low packing fractions, the authors report a linear force-velocity relation. At higher packing fraction, but {\it below} the glass transition a yield stress $F_0$ develops below which the probe remains trapped. For larger forces, the probe is delocalized by the applied force, the bead is pulled with a fluctuating velocity and a non-linear force-velocity relation holds. 
As emphasized by the authors themselves, the existence of a force threshold below the glass transition, where the spontaneous fluctuations alone still allow the particles to escapes their cages and relax, is rather intriguing.  Indeed, in a recent work~\cite{gazuz2009aan}, the force-velocity relation in dense suspensions has been investigated theoretically in the context of Mode Coupling Theory and compared to some of the above experimental data~\cite{habdas2004fmp} and to numerical simulations of a slightly polydisperse quasi-hard-sphere system undergoing strongly damped Newtonian dynamics. This time a force threshold is predicted to delocalize the probe particle {\it above} the glass transition. Below the glass transition a strongly non-linear force-velocity relation is predicted and reported in the simulations. This steep dependence of the velocity on the applied force could explain the above observation of a threshold in an experiment where the lower resolution on the velocity is always finite.

Recent simulations consider the motion of a single probe particle driven with a constant force in a binary mixture of two-dimensional disks with stiff spring repulsive interactions at zero temperature.~\cite{drocco2005mpj,reichhardt2009fjy}. As the packing fraction is increased towards jamming, the average velocity of the probe particle decreases and the velocity fluctuations show an increasingly intermittent or avalanche-like behavior. The velocity distributions are exponential away from jamming and have a power law character when approaching it within less than one percent. These observations are very similar to those reported in the present study. The averaged velocity dependence on the packing fraction exhibits the same faster than exponential decrease when approaching jamming and the power-law character of the velocity distributions close to jamming are identical. However the authors report the existence of a critical threshold force that must be applied for the probe particle to move through the sample, which increases when increasing the packing fraction {\it above} jamming, whereas we never observed a complete arrest of our probe.
In the simulations, once the probe stops, it cannot move any more because there is no temperature.
In the present experiments the vibration allows the system to explore new force chains configurations and thereby to provide the intruder new opportunities of moving. As a final remark on the force-velocity relation, we would like to emphasize that the affine relation, which we observe above threshold when the motion is continuous, is not to be confused with the linear response of Stoke's law. On the contrary, we believe that our observations correspond to the highest force regime reported in~\cite{gazuz2009aan}, where the non-linear response ultimately recovers a linear behavior.

An important difference between our system and the systems described above is that we are dealing with frictional particles. The role of friction close to jamming is an important issue. Recent important progress have been made in this matter~\cite{somfai2006can,song2008phase,henkes2009}, but many open questions remain unsolved. The isostatic criteria for mechanical stability $z=2d$ valid for frictionless systems \textit{a priori} turns into a double inequality $d+1<z<2d$ in the presence of friction, suggesting that friction could blur the critical character of jamming reported in frictionless systems. It was suggested in~\cite{reichhardt2009fjy} that the absence of power-law behavior in the velocity distribution observed in~\cite{geng2005sdi} might indeed be an effect of friction. Our observations clearly demonstrate on the contrary that the jamming transition remains critical in the presence of friction. Such a result suggests the existence of a generalized isostaticity criteria for frictional systems as proposed recently~\cite{henkes2009}. 

Not only the intruder motion exhibits the strongest intermittency at jamming, but also the statistics of the bursts obey critical scalings. The later are reminiscent of many other phenomena such as earthquakes~\cite{gutenberg1954sot}, Barkhausen noise~\cite{durin2006tbe}, crack tips dynamics in heterogeneous materials~\cite{maloy2006lwt}, generically assimilated to the so-called \textit{crackling noise}~\cite{sethna2001cn}. It is also a distinct behavior of various random failure and load redistribution models~\cite{hemmer2006mcc}, which can be used to describe stress redistribution in stick-slip granular experiments~\cite{dalton2005ssf}. These similarities, even at some quantitative level -- see in table~\ref{tab:comp} a comparison of some of the exponents obtained in different systems -- should not hide important differences among these phenomena even at the conceptual level. 
First the kind of criticality reported here has little to do with self organized criticality, often associated with crackling noise observations~\cite{sethna2001cn}. Second in many situations, an external parameter is increased (i.e. the loading force, the magnetic field) and the system fails once it has overcome some randomly distributed threshold. \textit{A contrario} in the present situation, the dragging force is kept constant, the system is vibrated and it explores successive configurations.
To what extent the study of these different dynamics are complementary is an interesting issue.
Further investigations with an intruder dragged at constant velocity should provide interesting clues in this direction.

\begin{table}[!t]
\begin{center}
\begin{tabular}[c]{|c|c|c|c|}
\hline
System [Ref.] & $1+\alpha$ & $1+\beta$ & $z$ \\ \hline
Here & $\sim3/2$ & $\sim1/3$ & $\sim2/3$\\
Barkhausen noise~\cite{durin2006tbe} & $1.5$ - $2.3$ & $1.24$ - $1.77$ & $1/2$ - $2/3$ \\
Crack tips~\cite{lockner1991qsf,diodati1991aef,cannelli1993soc,maes1998cic,guarino2002fta} & $1.2$ - $1.9$ & $1.47$ - $1.51$ & /\\
Stick skip (grains)~\cite{dalton2001soc,bretz2006bds,feder2005dss}& $1.8$ - $2.1$ & $1.67$ - $1.94$ & /\\ \hline
\end{tabular}
\end{center}
\caption{Comparison of the critical exponents for several systems displaying ``crackling noise''-like signals.}
\label{tab:comp}
\end{table}

Altogether, investigating the motion of a probe dragged at constant force in a dense granular media, we have identified a force threshold diverging at the jamming transition, below which the motion of the probe is intermittent and exhibits criticality at jamming.
More generally, we believe that micro and macro-rheological studies combined to statistical observations such as dynamical correlations are key elements to further investigate the underlying mechanisms of the jamming transition in frictional systems.



\begin{acknowledgements}
We would like to thank L. Ponson, E. Bouchaud, D. Bonamy, S. Aumaitre for helpful discussions, B. Dubrulle and M. Bonetti for clever advises on fractal dimensions computation, and F. Paradis, B. Saint-Yves and C. Coulais for their helpful hands when running the experiment and analysing the data. We also thank V. Padilla and C. Gasquet for technical assistance on the experiment. This work was supported by ANR DYNHET 07-BLAN-0157-01.
\end{acknowledgements}

\newpage 
\bibliography{biblio_glass}

\end{document}